\documentstyle[12pt]{article}
\def\beq{\begin{equation}}
\def\eeq{\end{equation}}
\def\bea{\begin{eqnarray}}
\def\eea{\end{eqnarray}}
\def\bq{\begin{quote}}
\def\eq{\end{quote}}

\def\AJ{{\it Astrophys.J.} }
\def\AJL{{\it Ap.J.Lett.} }
\def\AJS{{\it Ap.J.Supp.} }
\def\AM{{\it Ann.Math.} }
\def\AP{{\it Ann.Phys.} }
\def\APJ{{\it Ap.J.} }
\def\APP{{\it Acta Phys.Pol.} }
\def\ASAS{{\it Astron. and Astrophys.} }
\def\BAMS{{\it Bull.Am.Math.Soc.} }
\def\CMJ{{\it Czech.Math.J.} }
\def\CMP{{\it Commun.Math.Phys.} }
\def\FP{{\it Fortschr.Physik} }
\def\HPA{{\it Helv.Phys.Acta} }
\def\IJMP{{\it Int.J.Mod.Phys.} }
\def\JMM{{\it J.Math.Mech.} }
\def\JP{{\it J.Phys.} }
\def\JCP{{\it J.Chem.Phys.} }
\def\LNC{{\it Lett. Nuovo Cimento} }
\def\SNC{{\it Suppl. Nuovo Cimento} }
\def\MPL{{\it Mod.Phys.Lett.} }
\def\NAT{{\it Nature} }
\def\NC{{\it Nuovo Cimento} }
\def\NP{{\it Nucl.Phys.} }
\def\PL{{\it Phys.Lett.} }
\def\PR{{\it Phys.Rev.} }
\def\PRL{{\it Phys.Rev.Lett.} }
\def\PRTS{{\it Physics Reports} }
\def\PS{{\it Physica Scripta} }
\def\PTP{{\it Progr.Theor.Phys.} }
\def\RMPA{{\it Rev.Math.Pure Appl.} }
\def\RNC{{\it Rivista del Nuovo Cimento} }
\def\SJPN{{\it Soviet J.Part.Nucl.} }
\def\SP{{\it Soviet.Phys.} }
\def\TMF{{\it Teor.Mat.Fiz.} }
\def\TMP{{\it Theor.Math.Phys.} }
\def\YF{{\it Yadernaya Fizika} }
\def\ZETF{{\it Zh.Eksp.Teor.Fiz.} }
\def\ZP{{\it Z.Phys.} }
\def\ZMP{{\it Z.Math.Phys.} }

\def\gappeq{\mathrel{\rlap {\raise.5ex\hbox{$>$}}
{\lower.5ex\hbox{$\sim$}}}}

\def\lappeq{\mathrel{\rlap{\raise.5ex\hbox{$<$}}
{\lower.5ex\hbox{$\sim$}}}}

\begin{document}
\pagestyle{empty}
\begin{flushright}
{CERN-TH/99-365}
\end{flushright}
\vspace*{5mm}
\begin{center}
{\bf THE CRUCIAL PROBLEM:  THE ELECTROWEAK SYMMETRY BREAKING} \\
\vspace*{1cm} 
{ G. Altarelli}\\
\vspace{0.3cm}
Theoretical Physics Division, CERN \\
CH - 1211 Geneva 23 \\
and\\
Universit\`a di Roma Tre, Rome, Italy
\vspace*{2cm}
  
{\bf Contents} \\ 

\vspace*{5mm}
\noindent
\begin{itemize}
\item[1.] Why we do Believe in the Standard Model\\
\item[2.] Why we do not Believe in the Standard Model \\
\indent 2.1 Conceptual
Problems \\
\indent 2.2 Hints from Experiment \\
\indent~~~~~2.2.1 Unification of Couplings
\\
\indent~~~~~2.2.2 Dark Matter\\
\indent~~~~~2.2.3 Neutrino Masses
\\
\indent~~~~~2.2.4 Baryogenesis\\
\item[3.] Status of the Search for the Higgs and for New Physics\\
\item[4.] Conclusion\\
 \end{itemize}

\vspace*{2cm}

{\it Talk given at the XIth Rencontres de Blois\\
Frontiers of Matter\\
Blois, Loire Valley, France, 27 June -- 3 July 1999\\}
\end{center}
\vspace*{0.5cm}

\begin{flushleft} CERN-TH/99-365 \\
November 1999
\end{flushleft}
\vfill\eject

\setcounter{page}{1}
\pagestyle{plain}



\bibliographystyle{unsrt}    

\def\Journal#1#2#3#4{{#1} {\bf #2}, #3 (#4)}

\def\NCA{\em Nuovo Cimento}
\def\NIM{\em Nucl. Instrum. Methods}
\def\NIMA{{\em Nucl. Instrum. Methods} A}
\def\NPB{{\em Nucl. Phys.} B}
\def\PLB{{\em Phys. Lett.}  B}
\def\PRL{\em Phys. Rev. Lett.}
\def\PRD{{\em Phys. Rev.} D}
\def\ZPC{{\em Z. Phys.} C}

\def\st{\scriptstyle}
\def\sst{\scriptscriptstyle}
\def\mco{\multicolumn}
\def\epp{\epsilon^{\prime}}
\def\vep{\varepsilon}
\def\ra{\rightarrow}
\def\ppg{\pi^+\pi^-\gamma}
\def\vp{{\bf p}}
\def\ko{K^0}
\def\kb{\bar{K^0}}
\def\al{\alpha}
\def\ab{\bar{\alpha}}
\def\be{\begin{equation}}
\def\ee{\end{equation}}
\def\bea{\begin{eqnarray}}
\def\eea{\end{eqnarray}}
\def\CPbar{\hbox{{\rm CP}\hskip-1.80em{/}}}
\def\beq{\begin{equation}} 
\def\eeq{\end{equation}} 
\def\bea{\begin{eqnarray}} 
\def\eea{\end{eqnarray}}
\def\bq{\begin{quote}} 
\def\eq{\end{quote}}

\def\AJ{{\it Astrophys.J.} } 
\def\AJL{{\it Ap.J.Lett.} } 
\def\AJS{{\it Ap.J.Supp.} } 
\def\AM{{\it Ann.Math.} } 
\def\AP{{\it Ann.Phys.} } 
\def\APJ{{\it Ap.J.} } 
\def\APP{{\it Acta Phys.Pol.} }
\def\ASAS{{\it Astron. and Astrophys.} } 
\def\BAMS{{\it Bull.Am.Math.Soc.} } 
\def\CMJ{{\it Czech.Math.J.} } 
\def\CMP{{\it Commun.Math.Phys.} } 
\def\FP{{\it Fortschr.Physik} } 
\def\HPA{{\it Helv.Phys.Acta} } 
\def\IJMP{{\it Int.J.Mod.Phys.} } 
\def\JMM{{\it J.Math.Mech.} } 
\def\JP{{\it J.Phys.} } 
\def\JCP{{\it J.Chem.Phys.} } 
\def\LNC{{\it Lett. Nuovo Cimento} } 
\def\SNC{{\it Suppl. Nuovo Cimento} } 
\def\MPL{{\it Mod.Phys.Lett.} } 
\def\NAT{{\it Nature} } 
\def\NC{{\it Nuovo Cimento} }
\def\NP{{\it Nucl.Phys.} } 
\def\PL{{\it Phys.Lett.} } 
\def\PR{{\it Phys.Rev.} } 
\def\PRL{{\it Phys.Rev.Lett.} } 
\def\PRTS{{\it Physics Reports} }
\def\PS{{\it Physica Scripta} } 
\def\PTP{{\it Progr.Theor.Phys.} } 
\def\RMPA{{\it Rev.Math.Pure Appl.} } 
\def\RNC{{\it Rivista del Nuovo Cimento} }
\def\SJPN{{\it Soviet J.Part.Nucl.} } 
\def\SP{{\it Soviet.Phys.} } 
\def\TMF{{\it Teor.Mat.Fiz.} }
\def\TMP{{\it Theor.Math.Phys.} } 
\def\YF{{\it Yadernaya Fizika} } 
\def\ZETF{{\it Zh.Eksp.Teor.Fiz.} }
\def\ZP{{\it Z.Phys.} } 
\def\ZMP{{\it Z.Math.Phys.} }

\parskip 0.3cm

\def\gappeq{\mathrel{\rlap {\raise.5ex\hbox{$>$}} {\lower.5ex\hbox{$\sim$}}}}

\def\lappeq{\mathrel{\rlap{\raise.5ex\hbox{$<$}} {\lower.5ex\hbox{$\sim$}}}}


\section{Why we do Believe in the SM: Precision Tests}

In recent years new powerful tests of the Standard Model (SM) have been performed mainly at LEP but also
at SLC and at the Tevatron. The running of LEP1 was terminated in 1995 and close-to-final results of the data
analysis are now available~\cite{tim,ew}. The experiments at the
Z resonance have enormously improved
the accuracy of the data in the electroweak neutral current sector. The LEP2 programme is in progress and
will continue till the end of 2000. The top quark has been at last found at the Tevatron and the mass determined with few
percent accuracy. The errors on $m_Z$ and $\sin^2\theta_{eff}$ went down by two and one orders of magnitude respectively
since the start of LEP in 1989. Similar drastic progress has been made on $\alpha_s$, $m_W$ and the Higgs search.  The
validity of the SM has been confirmed to a level that we can say was unexpected. In the present data
there is no significant evidence for departures from the SM, no convincing hint of new physics. 
The impressive success of the
SM poses strong limitations on the possible forms of new physics. Favoured are models of the Higgs sector and of new physics
that preserve the SM structure  and only very delicately improve it, as is the case for fundamental Higgs(es) and
Supersymmetry. Disfavoured are models with a nearby strong non perturbative regime that  almost inevitably would affect the
radiative corrections, as for composite Higgs(es) or technicolour and its variants. 

The main lesson of the precision tests~\cite{sta} of the standard
electroweak theory can be summarised as follows. It has
been checked that the couplings of quark and leptons to the weak gauge bosons $W^{\pm}$ and $Z$ are indeed
precisely those prescribed by the gauge symmetry. The accuracy of a few $0.1\%$ for these tests implies that, not
only the tree level, but also the structure of quantum corrections has been verified. To a lesser accuracy the
triple gauge vertices
$\gamma W^+ W^-$ and
$Z W^+ W^-$ have also been found in agreement with the specific prediction, at the tree level, of the $SU(2)\bigotimes U(1)$
gauge theory. This means that it has been verified that the gauge symmetry is indeed unbroken in the
vertices of the theory: the currents are indeed conserved. Yet there is obvious evidence that the symmetry is
otherwise badly broken in the masses. In fact the $SU(2)\bigotimes U(1)$ gauge symmetry forbids masses for all
the particles that have been sofar observed: quarks, leptons and gauge bosons. But of all these particles
only the photon is massless (and the gluons protected by the $SU(3)$ colour gauge symmetry), all other are massive
(probably also the neutrinos). Thus the currents are conserved but the spectrum of particle states is not
symmetric. This is the definition of spontaneous symmetry breaking. The practical implementation of spontaneous
symmetry breaking in a gauge theory is via the Higgs mechanism. In the minimal SM one single
fundamental scalar Higgs isospin doublet is introduced and its vacuum expectation value v breaks the symmetry. All masses
are proportional to v, although the Yukawa couplings that multiply v in the
expression for the masses of quarks and leptons are distributed over a wide range. The Higgs sector is still very much
untested. The Higgs particle has not been found~\cite{janot} but its mass
can well be heavier than the present direct lower limit
$m_H\gappeq106$~GeV  from LEP2~\cite{9nov}~\footnote{In writing these
Proceedings, I include all information available in
November 1999.}.  One knew from the beginning that the Higgs search is
difficult: being coupled in proportion to masses one has
first to produce heavy particles and then try to detect the Higgs (itself heavy) in their couplings. What has been tested is
the relation
$m^2_W=m^2_Z
\cos^2{\theta_W}$, modified by computable radiative corrections. This relation means that the effective Higgs (be it
fundamental or composite) transforms indeed as a weak isospin doublet. 

Quantum
corrections to the electroweak precision tests depend on the masses and the couplings in the theory. For example they
depend on the top mass
$m_t$, the Higgs mass
$m_H$, the strong coupling $\alpha_s(m_Z)$, the QED coupling $\alpha(m_Z)$ (these are running couplings at the Z mass)
and other parameters which are better known. In particular quantum corrections depend quadratically on $m_t$ and only
logaritmically on $m_H$. From the observed radiative corrections one obtains a value of $m_t$ in fair agreement
with the observed value from the Tevatron. For the Higgs mass one finds a quantitative indication of the mass range
\cite{tim}: $\log_{10}{m_H({\rm GeV})}=1.88^{+0.28}_{-0.30}$ (or
$m_H=77^{+69}_{-39}$). This result on the Higgs mass is particularly remarkable. The value of
$\log_{10}{m_H({\rm GeV})}$ is right on top of the small window between
$\sim 2$ and $\sim 3$ which is allowed by the
direct limit, on the one side, and the theoretical upper limit on the Higgs mass in the minimal SM (see later),
$m_H\lappeq 600-800$~GeV, on the other side. If one had found a central
value like $\gappeq 4$ the model would have
been directly discarded. Thus the whole picture of a perturbative theory with a fundamental Higgs is well supported by the
data on radiative corrections. It is important that there is a clear indication for a particularly light Higgs. This is
quite encouraging for the ongoing search for the Higgs particle. More in general, if the Higgs couplings are removed
from the lagrangian the resulting theory is non renormalisable. A cutoff $\Lambda$ must be introduced. In the quantum
corrections 
$\log{m_H}$ is then replaced by $\log{\Lambda}$ plus a constant. The precise determination of the associated finite
terms would be lost (that is, the value of the mass in the denominator in the argument of the logarithm). Thus the fact
that, from experiment, one finds $\log{m_H}\sim 2$ is a strong argument in favour of the specific form of the Higgs
mechanism as in the SM. A heavy Higgs would need some unfortunate conspiracy 
\cite{ks}: the finite terms should accidentally
compensate for the heavy Higgs in the few key parameters of the
radiative corrections (e.g the $\epsilon$ parameters
\cite{sta})~\cite{bfs}. Or additional new physics,
 for example in the form of effective contact terms added to the minimal SM lagrangian,
should accidentally do the compensation~\cite{kh}, which again needs some
sort of conspiracy~\cite {bs}. 

\section{Why we do not Believe in the SM}
\subsection{Conceptual Problems}

	Given the striking success of the SM why are we not satisfied with that theory? Why not just find the Higgs
particle, for completeness, and declare that particle physics is closed? The main reason is that there are
strong conceptual indications for physics beyond the SM. 

	It is considered highly unplausible that the origin of the electro-weak symmetry breaking can be explained by
the standard Higgs mechanism, without accompanying new phenomena. New physics should be manifest at energies in
the TeV domain. This conclusion follows fron an extrapolation of the SM at very high energies. The computed
behaviour of the $SU(3)\otimes SU(2)\otimes U(1)$ couplings with energy clearly points towards the
unification of the electro-weak and strong forces (Grand Unified Theories: GUT's) at scales of energy
$M_{\rm GUT}\sim  10^{14}-10^{16}$~GeV~\cite{qqi} which are close to the
scale of
quantum gravity, $M_{\rm Pl}\sim 10^{19}$~GeV.
 One can also imagine  a unified theory of all interactions also including gravity (at
present superstrings~\cite{ler} provide the best attempt at such a
theory). Thus GUT's and the realm of quantum gravity set a
very distant energy horizon that modern particle theory cannot anymore ignore. Can the SM without new physics be
valid up to such large energies? This appears unlikely because the structure of the SM could not naturally
explain the relative smallness of the weak scale of mass, set by the Higgs mechanism at $\mu\sim
1/\sqrt{G_F}\sim  250$~GeV  with $G_F$ being the Fermi coupling constant.
 This so-called hierarchy problem~\cite{ssi} is related to the presence of
fundamental scalar fields in the theory with quadratic mass divergences
and no protective extra symmetry at $\mu=0$. For fermions, first, the divergences are logaritmic and, second, at
$m=0$ an additional symmetry, i.e. chiral  symmetry, is restored. Here, when talking of divergences we are not
worried of actual infinities. The theory is renormalisable and finite once the dependence on the cut off is
absorbed in a redefinition of masses and couplings. Rather the hierarchy problem is one of naturalness. If we
consider the cut off as a manifestation of new physics that will modify the theory at large energy scales, then it
is relevant to look at the dependence of physical quantities on the cut off and to demand that no unexplained
enormously accurate cancellations arise. 

	According to the above argument the observed value of 
$\mu\sim 250$~GeV is indicative of the existence of new
physics nearby. There are two main possibilities. Either there exist fundamental scalar Higgses but the theory
is stabilised by supersymmetry, the boson-fermion symmetry, that would downgrade the degree of divergence from
quadratic to logarithmic. For approximate supersymmetry the cut off is replaced by the splitting between the
normal particles and their supersymmetric partners. Then naturalness demands that this splitting (times the
size of the weak gauge coupling) is of the order of the weak scale of mass, i.e. the separation within
supermultiplets should be of the order of no more than a few TeV. In this case the masses of most supersymmetric
partners of the known particles, a very large managerie of states, would fall, at least in part, in the discovery
reach of the LHC. There are consistent, fully formulated field theories constructed on the basis of this idea, the
simplest one being the MSSM~\cite{43}. As already mentioned, all normal
observed states are those whose masses are
forbidden in the limit of exact
$SU(2)\otimes U(1)$. Instead for all SUSY partners the masses are allowed in that limit. Thus when
supersymmetry is broken in the TeV range but $SU(2)\otimes U(1)$ is intact only s-partners take mass while all
normal particles remain massless. Only at the lower weak scale the masses of ordinary particles are generated.
Thus a simple criterium exists to understand the difference between particles and s-particles.

	The other main avenue is compositeness of some sort. The Higgs boson is not elementary but either a bound
state of fermions or a condensate, due to a new strong force, much stronger than the usual strong interactions,
responsible for the attraction. A plethora of new "hadrons", bound by the new strong force would  exist in the
LHC range. A serious problem for this idea is that nobody sofar has been  able to build up a realistic model
along these lines, but that could eventually be explained by a lack of ingenuity on the theorists side. The
most appealing examples are technicolour
theories~\cite{30,chi}. These models were inspired by the
breaking of chiral symmetry in massless QCD induced by quark condensates. In the case of the electroweak
breaking new heavy techniquarks must be introduced and the scale analogous to $\Lambda_{QCD}$ must be about
three orders of magnitude larger. The presence of such a large force relatively nearby has a strong tendency to
clash with the results of the electroweak precision tests~\cite{32}. New
versions have been developed~\cite{chi} to overcome the
negative response of the data,
 but models are far from offering a realistic picture.

Are there other ways to solve the hierarchy problem?
 Recently an exotic way was proposed~\cite{extra,giud}. The
idea is
that perhaps the scale of gravity is only apparently so large. It has been shown that it is in principle possible to bring
down the scale of gravity in the multi TeV energy range. This can happen if one assumes the existence of extra space
dimensions with sufficiently large compactification radius, with the graviton propagating in all dimensions, while ordinary
gauge interactions are trapped on a four dimensional wall. The corresponding modification of gravity at submillimetric
distances is compatible with existing limits. The vicinity of the decompactification scale can manifest itself in high energy
processes at $e^+e^-$ and hadron colliders where gravitons can be produced and appear as missing energy. This very
speculative scenario is certainly interesting especially as a stimulus to look for specific signals. But does not appear
as particularly compelling because the reason why the 
decompactification scale should be $\simeq$~few~TeV remains
mysterious. In addition all the positive hints we have in favour of the ordinary picture of GUTs from coupling unification,
neutrino masses, dark matter and so on would be emptied. Finally early time cosmology should be rewritten.  

The hierarchy problem is certainly not the only conceptual problem of the SM. There are many more: the
proliferation of parameters, the mysterious pattern of fermion masses and so on. But while most of these
problems can be postponed to the final theory that will take over at very large energies, of order
$M_{\rm GUT}$ or
$M_{\rm Pl}$, the hierarchy problem arises from the unstability of the low energy theory and requires a solution at
relatively low energies. 

A supersymmetric extension of the SM provides a way out which is well defined,
computable and that preserves all virtues of the SM.  The necessary
SUSY breaking~\cite{mura} can be introduced through soft
terms that do not spoil the good convergence properties of the theory. Precisely those terms arise from
supergravity when it is spontaneoulsly broken in a hidden sector. This is the case in the Minimal
Supersymmetric Standard Model (MSSM)~\cite{43}.   In this
most traditional approach SUSY is broken in a hidden sector~\cite{yyi} and
the scale of SUSY breaking is very
large of order
$\Lambda\sim\sqrt{G^{-1/2}_F M_{\rm Pl}}$  where
$M_{\rm Pl}$ is the Planck mass. But since the hidden sector only communicates with the visible sector
through gravitational interactions the splitting of the SUSY multiplets is much smaller, in the TeV
energy domain, and the Goldstino is practically decoupled. 
But alternative mechanisms of SUSY breaking are also
 being considered
\cite{giud,gauge,anom}. 
In one alternative scenario the (not so
much) hidden sector is connected to the visible one by ordinary gauge interactions. As these are much
stronger than the gravitational interactions, $\Lambda$ can be much smaller, as low as 
10-100~TeV. It follows that the Goldstino is very light in these models (with mass of
order or below 1~eV
typically) and is the lightest, stable SUSY particle, but its couplings are observably large. The radiative
decay of the lightest neutralino into the Goldstino leads to detectable photons. The signature of photons comes
out naturally in this SUSY breaking pattern: with respect to the MSSM, in the gauge mediated model there are typically
more photons and less missing energy. The main appeal of gauge mediated models is a better protection against
flavour changing neutral currents. In the gravitational version even if we accept that gravity leads to
degenerate scalar masses at a scale near $M_{\rm Pl}$ the running of the masses down to the weak scale can
generate mixing induced by the large masses of the third generation
 fermions~\cite{giud}. More recently it has been
pointed out~\cite{anom} that there are pure gravity contributions to soft
masses that arise from gravity theory
anomalies. In the assumption that these terms are dominant the associated spectrum and phenomenology has been studied. In
this case gaugino masses are proportional to gauge coupling beta functions, so that the gluino is much heavier than the
electroweak gauginos, and the wino is most often the lightest SUSY
particle. 

The MSSM~\cite{43} is a completely specified,
consistent and computable theory. There are too many parameters to attempt a direct fit of the data to
the most general framework. But we can consider two significant limiting cases: the "heavy" and the
"light" MSSM.

	The "heavy" limit corresponds to all s-particles being sufficiently massive, still within the limits
of a natural explanation of the weak scale of mass. 
In this limit a very important result holds~\cite{58}: 
for what concerns the precision electroweak tests, the MSSM predictions tend to reproduce
the results of the SM with a light Higgs, say $m_H\sim$ 100~GeV. So if the
masses of SUSY partners are pushed
at sufficiently large values the same quality of fit as for the SM is guaranteed. 

	In the "light" MSSM option some of the superpartners have a relatively small mass, close to their
experimental lower bounds. In this case the pattern of radiative corrections may sizeably deviate from
that of the SM~\cite{pok}. The potentially largest effects occur in vacuum
polarisation amplitudes and/or the
$Z\rightarrow b\bar b$ vertex. Since no sign of deviations from the SM is seen in the data and no light SUSY partners
have been found at LEP2 or at the Tevatron, the "light" case can no more be that light.

According to the prevailing view at present, the large scale structure of particle physics consists of a unified theory
at
$M=M_{\rm GUT}-M_{\rm Pl}$ and a low energy effective theory valid at and above the weak scale of energy.  The lagrangian density 
of the low energy  effective theory, after integrating out all very heavy degrees of
freedom,  consists of a set of operators of dimension non larger than 4, that correspond to the renormalisable part, plus a
set of higher dimension, non renormalisable, operators. Schematically, we have:
\beq
{\cal L}=\mu^2 \phi^2+m \bar\psi \psi + g \bar\psi iD\llap{$/$} \psi+\lambda \phi^4+......+
\frac{\lambda_5}{M}\bar\psi \psi \phi \phi+\frac{\lambda_6}{M^2}\bar\psi \psi \bar\psi \psi+....\label{eff}
\eeq
Indicatively, we have shown a number of typical terms of dimension 2 (boson masses), 3 (fermion masses), 4
(renormalisable interactions) plus examples of operators of higher
dimension, 5 and 6. Due to the very
large scale of energy where the really fundamental theory applies, the conditions on the low energy effective theory
are severe. First, the dimension $\leq4$ part must be renormalisable. This is a minimum requirement in order to have a closed,
consistent and predictive description of the dynamics after the presence of the very high cut off has been hidden inside
renormalised masses and couplings. But this is not enough because the dependence of masses and couplings from the cut off must
be reasonable in order to avoid the necessity of immense fine tuning. For this to be true additional conditions must be
satisfied. The coupling in front of each
operator, in absence of specific reasons, should be proportional to the large cut off $M$ raised to a power d fixed by
dimensions. For example, $\mu^2$ should be proportional to $M^2$. In the SM there is no symmetry reason why this should
not be the case. So boson masses, like the W and Z masses, should be of
order M. This the hierarchy problem~\cite{ssi} . In
supersymmetric extensions of the SM $\mu^2$ is instead of order the mass splittings of SUSY multiplets, because in the
limit of exact SUSY symmetry there are no quadratic divergences (in presence of boson-fermion symmetry the stronger
bosonic divergences must disappear, in order that bosonic and fermionic divergences can both be logaritmic). For fermions
$m$ is not of order $M$ but of order
$v\log{M}$ because the divergences in the fermionic sector are always at most logaritmic. Also, chiral symmetry ensures
that if you start from zero masses the quantum corrections to $m$ must vanish. Once supersymmetry
or some other stabilising mechanism is introduced, the renormalisable part of the lagrangian is sufficiently insensitive
to the presence of the very large cut off $M$. The additional non renormalisable terms are suppressed by powers of $M$.
At energies of order $v$, the electro-weak scale, their effects are proportional to $(v/M)^d$, $d=1,2,...$, hence very
small.

\subsection{Hints from Experiment}
\subsubsection{Unification of Couplings}

At present the most direct
phenomenological evidence in favour of supersymmetry is obtained from the unification of couplings in GUTs.
Precise LEP data on $\alpha_s(m_Z)$ and $\sin^2{\theta_W}$ confirm what was already known with less accuracy:
standard one-scale GUTs fail in predicting $\sin^2{\theta_W}$ given
$\alpha_s(m_Z)$ (and $\alpha(m_Z)$) while SUSY GUTs~\cite{zzi} are in
agreement with the present, very precise,
experimental results. According to the analysis of ref.~\cite{aaii}, if
one starts from the known values of
$\sin^2{\theta_W}$ and $\alpha(m_Z)$, one finds for $\alpha_s(m_Z)$ the results:
\bea
		\alpha_s(m_Z) = 0.073\pm 0.002 ~~~~~      	(\rm{Standard~ GUTs})\nonumber \\	
		\alpha_s(m_Z) = 0.129\pm0.010~~~~~  (\rm{SUSY~ GUTs})
\label{24}
\eea
to be compared with the world average experimental value $\alpha_s(m_Z)$ =0.119(4).

\subsubsection{Dark Matter}

There is solid astrophysical and cosmological 
evidence~\cite{kol,spi} that most of the matter in the
universe
does not emit electromagnetic radiation, hence is "dark". Some of the dark matter must be baryonic but most of it must
be non baryonic. Non baryonic dark matter can be cold or hot. Cold means non relativistic at freeze out, while hot is
relativistic. There is general consensus that most of the non baryonic dark matter must be cold dark matter. A couple
of years ago the most likely composition was quoted to be around $80\%$ cold and $20\%$ hot. At present it appears
that the need of a sizeable hot dark matter component is more uncertain. In fact, recent experiments have indicated the
presence of a previously disfavoured cosmological constant component in
$\Omega=\Omega_m+\Omega_{\Lambda}$~\cite{kol}. Here
$\Omega$ is the total matter-energy density in units of the critical density, $\Omega_m$ is the matter component
(dominated by cold dark matter) and $\Omega_{\Lambda}$ is the cosmological component. Inflationary theories strongly
favour
$\Omega=1$ which is consistent with present data. At present, still within large uncertainties, the approximate
composition is indicated to be
$\Omega_m\sim 0.4$ and
$\Omega_{\Lambda}\sim0.6$ (baryonic dark matter gives $\Omega_b\sim0.05$). 

The implications for particle physics is that certainly there must exist a source of cold dark matter. By far the
most appealing candidate is the neutralino, the lowest supersymmetric particle, in general a superposition of
photino, Z-ino and higgsinos. This is stable in supersymmetric models with R parity conservation, which are the
most standard variety for this class of models (including the MSSM). A
neutralino with mass of order 100~GeV would fit perfectly as a cold dark
matter candidate. Another common
candidate for cold dark matter is the axion, the elusive particle associated to a possible solution of the strong
CP problem along the line of a spontaneously broken Peccei-Quinn symmetry. To my knowledge and taste this
option is less plausible than the neutralino. One favours supersymmetry for very diverse conceptual and
phenomenological reasons, as described in the previous sections, so that neutralinos are sort of standard by now.
For hot dark matter, the self imposing candidates are neutrinos. If we demand a density fraction
$\Omega_{\nu}\sim 0.1$ from neutrinos, then it turns out that the sum of stable neutrino masses should be around 5
eV~\cite{kol}.

\subsubsection{Neutrino Masses}

Recent data from Superkamiokande~\cite{SK,WIN99} have
provided a more solid experimental basis for neutrino
oscillations as an explanation of the atmospheric neutrino anomaly. In
addition the solar neutrino deficit~\cite{bel},
observed by several experiments, is also probably an indication of a different sort of neutrino oscillations. Results
from the laboratory experiment by the LSND
collaboration~\cite{acc,LSND} can also be considered as a
possible indication of
yet another type of neutrino oscillation.  Neutrino oscillations imply neutrino masses. The extreme smallness of neutrino
masses in comparison with quark and charged lepton masses indicate a different nature of neutrino masses, linked to
lepton number violation and the Majorana nature of neutrinos. Thus neutrino masses provide a window on the very large
energy scale where lepton number is violated and on GUTs. The new experimental evidence on
neutrino masses could also give an important feedback on the problem of quark and charged lepton masses, as all these
masses are possibly related in GUTs. In particular the observation of a nearly maximal mixing angle for
$\nu_{\mu}\rightarrow \nu_{\tau}$ is particularly interesting. Perhaps also solar neutrinos may occur with
large mixing angle. At present solar neutrino mixings can be either large or very small, depending on which particular
solution will eventually be established by the data. Large mixings are very interesting because a first guess was in
favour of small mixings in the neutrino sector in analogy to what is observed for quarks. If confirmed, single or double
maximal mixings can provide an important hint on the mechanisms that generate neutrino masses.

The experimental status of neutrino oscillations is still very preliminary . While the evidence for the
existence of neutrino oscillations from solar and atmospheric neutrino data is rather convincing by now, the values of
the mass squared differences $\Delta m^2$ and mixing angles are not firmly established. For solar neutrinos, for example,
three possible solutions are still possible~\cite{solar}. Two are based on
the MSW mechanism~\cite{MSW}, one with small
(MSW-SA:
$\sin^2{2\theta_{sun}}\sim 5.5~10^{-3}$) and one with large mixing angle (MSW-LA:  $\sin^2{2\theta_{sun}}\gappeq 0.2$),
and one in terms of vacuum oscillations (VO) with large mixing angle (VO:
$\sin^2{2\theta_{sun}}\sim 0.75$). For atmospheric neutrinos the preferred value of
$\Delta m^2$ is affected by large uncertainties and could still sizeably drift in one sense or the other, but the fact
that the mixing angle is large appears established ($\sin^2{2\theta_{atm}}\gappeq 0.9 ~{\rm at}~ 90\%~ {\rm C.L.}$)
\cite{fogli,hall,WIN99}. Another issue which is still open is
the claim by the LSND collaboration of an additional 
signal of
neutrino oscillations in a reactor experiment \cite{acc}. This claim was not so-far supported by a second recent
experiment, Karmen~\cite{Karmen}, but the issue is far from being closed.
Given the present experimental uncertainties the
theorist has to make some assumptions on how the data will finally look like in the future. Here we tentatively assume
that the LSND evidence will disappear. If so then we only  have two oscillations frequencies, which can be given in terms
of the three known species of light neutrinos without additional sterile kinds (i.e. without weak interactions, so that
they are not excluded by LEP). We then take for granted that the frequency of atmospheric neutrino oscillations will remain
well separated from the solar neutrino frequency, even for the MSW
solutions. The present best values are~\cite{solar,fogli,hall,WIN99} 
$(\Delta m^2)_{atm}\sim 3.5~10^{-3}~{\rm eV}^2$ and $(\Delta m^2)_{MSW-SA}\sim
5~10^{-6}~{\rm eV}^2$ or
$(\Delta m^2)_{VO}\sim 10^{-10}~{\rm eV}^2$. We also assume that the electron neutrino does not participate in the atmospheric
oscillations, which (in absence of sterile neutrinos) are interpreted as nearly maximal
$\nu_{\mu}\rightarrow\nu_{\tau}$ oscillations as indicated by the
Superkamiokande~\cite{SK,WIN99} and Chooz~\cite{Chooz} data. However the
data do not exclude a non-vanishing $U_{e3}$ element.
 In the Superkamiokande allowed
region the bound by Chooz~\cite{Chooz} amounts to  $|U_{e3}|\lappeq 0.2$
\cite{fogli,hall}.

In summary, by now it is very unlikely that all this
evidence for neutrino oscillations will disappear or be explained away by astrophysics or other solutions. The
consequence is that we have a substantial evidence that neutrinos are
massive. From a strict minimal standard model point
of view neutrino masses could vanish if no right handed neutrinos existed (no Dirac mass) and lepton number was
conserved (no Majorana mass). In GUTs both these assumptions are violated. The right handed neutrino is required in all
unifying groups larger than SU(5). In SO(10) the 16 fermion fields in each family, including the right handed neutrino,
exactly fit into the 16 dimensional representation of this group. This is really telling us that there is something in
SO(10)! The SU(5) alternative in terms of $\bar 5+10$, without a right handed neutrino, is certainly less elegant. The
breaking of
$|B-L|$, B and L is also a generic feature of GUTs. In fact, the see-saw 
mechanism~\cite{ssm} explains
the smallness of neutrino masses in terms of the large mass scale where $|B-L|$ and L are violated. Thus, neutrino
masses, as would be proton decay, are important as a probe into the physics at the GUT scale.

Oscillations only determine squared mass differences and not masses. The case of three nearly degenerate neutrinos
is the only one that could in principle accomodate neutrinos as hot dark matter together with solar and atmospheric
neutrino oscillations. According to our previous discussion, the common mass
should be around 1-3~eV. The solar
frequency could be given by a small 1-2 splitting, while the atmospheric frequency could be given by a still small
but much larger 1,2-3 splitting. A strong constraint arises in the degenerate case from neutrinoless double beta
decay which requires that the ee entry of
$m_{\nu}$ must obey
$|(m_{\nu})_{11}|\leq 0.2-0.5~{\rm eV}$~\cite{dbeta}. As observed in
ref.~\cite{GG}, this bound can only be 
satisfied if
double maximal mixing is realized, i.e. if also solar neutrino oscillations occur with nearly maximal mixing.
We have mentioned that it is not at all clear at the moment that a hot dark matter component is really
needed~\cite{kol}. However the only reason to consider the fully
degenerate solution is 
that it is compatible
with hot dark matter.
Note that for degenerate masses with $m\sim 1-3~{\rm eV}$ we need a relative splitting $\Delta m/m\sim
\Delta m^2_{atm}/2m^2\sim 10^{-3}-10^{-4}$ and an even smaller one for solar neutrinos. It is not simple
to imagine a natural mechanism compatible with unification and the see-saw mechanism to arrange such a
precise near symmetry.

If neutrino masses are smaller than for cosmological relevance, we can have the hierarchies $|m_3| >> |m_{2,1}|$
or $|m_1|\sim |m_2| >> |m_3|$. Note that we
are assuming only two frequencies, given by $\Delta_{sun}\propto m^2_2-m^2_1$ and
$\Delta_{atm}\propto m^2_3-m^2_{1,2}$. We prefer the first case, because for quarks and leptons one
mass eigenvalue, the third generation one, is largely dominant. Thus the dominance of $m_3$ for neutrinos
corresponds to what we observe for the other fermions.  In this case, $m_3$ is determined by the atmospheric
neutrino oscillation frequency to be around $m_3\sim0.05$~eV. By the see-saw mechanism $m_3$ is related to some
large mass M, by $m_3\sim m^2/M$. If we identify m with either the Higgs vacuum expectation value or the top mass
(which are of the same order), as suggested for third generation neutrinos by GUTs in simple SO(10)
models, then M turns out to be around $M\sim 10^{15}$~GeV, which is
consistent with the connection with GUTs. If
solar neutrino oscillations are determined by vacuum oscillations, then $m_2\sim
10^{-5}$~eV and we have that the
ratio $m_2/m_3$ is well consistent with $(m_c/m_t)^2$.

A lot of attention~\cite{us} is being devoted to the
problem of a natural explanation of the observed nearly maximal mixing angle for atmospheric
neutrino oscillations and possibly also for solar neutrino oscillations, if explained by vacuum
oscillations. Large mixing angles are somewhat unexpected because
the observed quark mixings are small and the quark, charged lepton and neutrino mass matrices are to
some extent related in GUT's. There must be some special interplay between the neutrino Dirac
and Majorana matrices in the see-saw mechanism in order to generate maximal
mixing. It is hoped that looking for a natural explanation of large neutrino mixings can lead us to decripting
some interesting message on the physics at the GUT scale.

\subsubsection{Baryogenesis}

Baryogenesis is interesting because it could occur at the weak
scale~\cite{rub} but not in the SM. For baryogenesis one needs the three
famous Sakharov conditions~\cite{sak}: B
violation, CP violation and no termal equilibrium. In principle these conditions could be verified in the SM. B is
violated by instantons when kT is of the order of the weak scale (but B-L is conserved). CP is violated by the CKM
phase and out of equilibrium conditions could be verified during the electroweak phase transition. So the
conditions for baryogenesis  at the weak scale in the SM appear superficially to be present.
However, a more quantitative analysis~\cite{rev,cw1} shows that
baryogenesis is not possible
in the SM because there is not enough CP violation and the phase transition is not sufficiently strong first order,
unless
$m_H<80$~GeV, which is by now excluded by LEP. However, it is interesting
that baryogenesis at the weak scale is not yet
excluded in SUSY extensions of the SM
\cite{cw1}. In particular, in the MSSM there are additional sources of CP violations and the bound on $m_H$ is modified
by a sufficient amount by the presence of scalars with large couplings to the Higgs sector, typically the s-top. What is
required is that
$m_h\sim 80-110$~GeV, a s-top not heavier than the top quark and,
preferentially, a small
$\tan{\beta}$. This possibility is becoming more and more marginal with the progress of the LEP2 running and will be
completely excluded if no signals are found in last phase of LEP2 operation. 

If baryogenesis at the weak scale is excluded by the data it can occur at or just below the
GUT scale, after inflation. But only that part with
$|B-L|>0$ would survive and not be erased at the weak scale by instanton effects. Thus baryogenesis at $kT\sim
10^{12}-10^{15}$~GeV needs B-L violation at some stage like for $m_\nu$,
if neutrinos are Majorana particles. The two
effects could be related if baryogenesis arises from
leptogenesis~\cite{lg} then converted into baryogenesis by
instantons. Recent results on neutrino masses are compatible with this
possibility~\cite{buch}. Thus the possibility of
baryogenesis at a large energy scale has been  boosted by the recent results on neutrinos.

\section{Status of the Search for the Higgs and for New Physics}

The LEP2 programme has started in the second part of 1995. At first the energy was fixed at 161~GeV, which is
the most favourable energy for the measurement of $m_W$ from the cross-section for
$e^+e^- \rightarrow W^+W^-$ at threshold. Then gradually the energy was
brought up to 172, 183, 189~GeV. In ' 99 it was
increased up to a maximum of 202~GeV with a record integrated
luminosity in one year of $254 pb^{-1}$~~\cite{9nov}. LEP2 will resume
the run in spring 2000, increasing the energy by a few more GeV, before its dismantlement at  the end of 2000 for the
installation of the LHC ring in the tunnel. The main goals of LEP2 are the search for the Higgs and for new particles,
the measurement of
$m_W$ and the investigation of the triple gauge vertices
$WWZ$ and $WW\gamma$.  
A complete survey of the LEP2 
physics is collected in the two volumes of ref.~\cite{lep2}. 

	An important competitor of LEP2 is the Tevatron collider. In 2000-01 the Tevatron will start RunII with the purpose
of collecting a few $fb^{-1}$ of integrated luminosity at $2$~TeV. The competition is especially on the search of
new particles, but also on
$m_W$ and the triple gauge vertices. For example, for supersymmetry while the Tevatron is superior for gluinos and
squarks,  LEP2 is strong on Higgses, charginos, neutralinos and sleptons. There are plans for RunIII to start in
$\gappeq 2004$ with the purpose
of collecting of the order  $5~fb^{-1}$ of integrated luminosity per year. If so the Tevatron could also hope to find
the Higgs before the LHC if the Higgs mass is close to the LEP2 range.

Concerning the Higgs, the present limits obtained by the LEP collaborations at the end of the '99 run and still
preliminary and not combined, are, for the SM Higgs, $m_H\gappeq 106$~GeV
and for the lightest MSSM Higgs, $m_h\gappeq 90$~GeV~\cite{9nov}. To
understand the significance of these limits we recall
the theoretical bounds on the Higgs mass. 
	
It is well known~\cite{zziii}$^-$\cite{cciiii} that in the SM with only one
Higgs doublet a lower limit on
$m_H$ can be derived from the requirement of vacuum stability. This criterium is equivalent to demand that the coupling
$\lambda$ of the quartic term $\lambda (\phi \dagger \phi)^2$ does not become negative while running from the weak
scale up to the scale $\Lambda$. The initial value of $\lambda$ at the weak scale increases with $m_H^2$, while the
derivative, for $m_H$ near the limit, is dominated by the top quark term which is large and negative. The value of the limit is
a function of
$m_t$ and of the energy scale
$\Lambda$ where the model breaks down and new physics appears.  
If one requires that
$\lambda$ remains positive up to $\Lambda = 10^{15}$--$10^{19}$~GeV, then
the resulting bound on $m_H$ in the SM with
only one Higgs doublet is given by~\cite{aaiiii}:
\begin{equation} m_H > 134 + 2.1 \left[ m_t - 173.8 \right] - 4.5~\frac{\alpha_s(m_Z) - 0.119}{0.006}~.
\label{25h}
\end{equation}

We see that the discovery of a Higgs particle at
LEP2, or $m_H\lappeq 110$~GeV, would imply that the SM breaks down at a
scale
$\Lambda$ of the order of $\lappeq 100$~TeV. It can be shown~\cite{zziii}
that the lower limit is not much relaxed even
if strict vacuum stability is replaced by some sufficiently long metastability.

Similarly an upper bound on $m_H$ (with mild dependence
on $m_t$) is obtained \cite{eeiiii} from the requirement that up to the scale $\Lambda$ no Landau pole appears. The upper limit
on the Higgs mass in the SM is important to guarantee the success of the LHC as an accelerator designed to solve
the Higgs problem.  In
fact, for large Higgs masses, the initial value of $\lambda$ is large and the derivative of $\lambda$ is positive, because
the positive $\lambda$ term (the $\lambda \phi^4$ theory is not asymptotically free!) overwhelms the top Yukawa negative
contribution. 
As a consequence the coupling $\lambda$ tends to 
infinity (the Landau pole) at some finite scale. 
The upper limit on $m_H$ has been recently reevaluated~\cite{hr}. For
$m_t\sim 175$~GeV one finds
$m_H\lappeq 180$~GeV for $\Lambda\sim M_{GUT}-M_{Pl}$ and $m_H\lappeq
0.5-0.8$~TeV for $\Lambda\sim
1$~TeV. Actually, for
$m_t \sim$ 174~GeV, only a small range of values for $m_H$ is allowed, 
$130 < m_H <~\sim 200$~GeV, if the SM holds up
to $\Lambda \sim M_{GUT}$ or $M_{\rm Pl}$~\cite{hr}. 

A particularly
important example of theory where the above bounds do not apply and in particular the lower bound is violated, is the
MSSM, which we now discuss. As is well known~\cite{43}, in the MSSM there
are two Higgs doublets, which implies three
neutral physical Higgs particles and a pair of charged Higgses. The lightest neutral Higgs, called $h$, should be
lighter than
$m_Z$ at tree-level approximation. However, radiative
corrections~\cite{ffiiii} increase the $h$ mass by a term
proportional to $m^4_t$ and  logaritmically dependent on the stop mass . Once the radiative corrections are taken into
account the $h$ mass still remains rather small: for $m_t = 174$~GeV one
finds the limit $m_h \lappeq 130$~GeV 
(valid for all values of $tg\beta$ and saturated at large 
$tg\beta$)~\cite{ddiiii}. Actually one can well expect that $m_h$ is
sizeably
below the bound if $tg\beta$ is small. LEP is now progressively
eliminating the small  $tg\beta$ region~\cite{9nov}.

Another main goal of LEP2 is the search for direct signals of supersymmetry. By now most of
the discovery potential of LEP2 for supersymmetry has been deployed. For example, the limit on the chargino mass
was
about $m_{\chi^+}\gappeq 45$~GeV after LEP1 and is now about
$m_{\chi^+}\gappeq 100$~GeV, apart from exceptional regions of the MSSM
parameter
space. The lightest neutralino mass limit is around
$m_{\chi^0}\gappeq36$~GeV~\cite{9nov}. The region of the MSSM parameter
space that
has been by now excluded by LEP is a very important one. The low
$tg\beta$ solution was appealing in many respects.
 With no discovery of the Higgs and SUSY at LEP the case for the MSSM becomes
less natural, and even less natural become the gauge mediated models , 
in the sense of, for example, refs.~\cite{nat}. Similarly, some more
constrained forms of the model, like the supergravity version, where degenerate scalar masses and gaugino masses are assumed
at the GUT scale, are by now disfavoured becoming increasingly
unnatural.But naturaleness is not a completely quantitative criterium
so that the issue is open until the upper bound  $m_H\lappeq130$~GeV
for the general MSSM is not disproven.

\section{Conclusion}

Today in particle physics we follow a double approach: from above and from below. From above there are, on the theory
side, quantum gravity (that is superstrings), GUT theories and cosmological scenarios. On the experimental side there
are underground experiments (e.g. searches for neutrino oscillations and proton decay), cosmic ray
observations, satellite experiments (like COBE, IRAS etc) and so on. From below, the main objectives of theory and
experiment are the search of the Higgs and of signals of particles beyond the Standard Model (typically supersymmetric
particles). Another important direction of research is aimed at the exploration of the flavour problem: study of CP
violation and rare decays. The general expectation is that new physics is close by and that should, be found very
soon if not for the complexity of the necessary experimental technology that makes the involved time scale painfully
long.


\end{document}